\documentclass[12pt,aps,prb]{revtex4}   

\usepackage{amsmath}    
\usepackage{graphicx}   
\usepackage{setspace} 

\begin{document}

\title{Optimal Pacing for Running 400 m and 800 m Track Races}
\author{James Reardon}
 \affiliation{University of Wisconsin--Madison, Department of Physics, Madison,
WI 53706}
 \email{reardon@physics.wisc.edu}   
\date{\today}

\singlespacing

\begin{abstract}
Physicists seeking to understand complex biological systems often find it rewarding to create simple ``toy models" that reproduce system behavior.  Here a toy model is used to understand a puzzling phenomenon from the sport of track and field.  Races are almost always won, and records set, in 400 m and 800 m running events by people who run the first half of the race faster than the second half, which is not true of shorter races, nor of longer.  There is general agreement that performance in the 400 m and 800 m is limited somehow by the amount of anaerobic metabolism that can be tolerated in the working muscles in the legs.  A toy model of anaerobic metabolism is presented, from which an optimal pacing strategy is analytically calculated via the Euler-Lagrange equation.  This optimal strategy is then modified to account for the fact that the runner starts the race from rest; this modification is shown to result in the best possible outcome by use of an elementary variational technique that supplements what is found in undergraduate textbooks.  The toy model reproduces the pacing strategies of elite 400 m and 800 m runners better than existing models do.  The toy model also gives some insight into training strategies that improve performance.     
\end{abstract}

\maketitle
\newpage

\section{Introduction}
The sport of athletics, called ``track and field" in the USA, includes running competitions at distances ranging from 60 m to 10000 m.  Accurate records of finish times exist, going back more than a century, of a multitude of races at a variety of standard distances, including ``world records", in which an athlete ran faster for a given race distance than anyone had run before.  For a many of these world-record performances, accurate measurements were made of the time it took the record-breaking athlete to reach the halfway point of the race.  The subset of world-record performances for standard track races, as ratified by the International Association of Athletics Federations (IAAF), for which halfway ``split" times have been published by the IAAF,\cite{IAAF} is summarized in Table~\ref{tab:ld_data}.  
Here ``one-lap" includes the 400 m and its imperial equivalent, the 440 yd=402.34 m, and ``two-lap" includes the 800 m and 880 yd=804.67 m.  All world records for the 100 m and 200 m have been set with ``negative splits", in which the second half of the race distance is covered faster than the first half.  All world records for one-lap races have been set with ``positive splits" (second half slower than first half).  Almost all records for two-lap races have been set with positive splits.  For longer races there is no strong preference for positive splits or negative splits.  What makes the one-lap and two-lap races different from shorter and longer races?
\begin{table}[h!]
\begin{center}
\begin{tabular}{|c|c|c|c|c|c|}
\hline
 & \multicolumn{2}{c|}{men} & \multicolumn{2}{c|}{women} \\
\hline
 & positive split & negative split & positive split & negative split \\
 \hline
 100m & 0 & 6 & 0 & 0 \\
 \hline
 200m & 0 & 6 & 0 & 3 \\
 \hline
 one-lap & 25 & 0 & 9 & 0 \\
\hline
two-lap & 27 & 2 & 13 & 1 \\
\hline
Mile & 19 & 12 & 5 & 4 \\
\hline
3000m & 12 & 9 & 0 & 2 \\
\hline
10000m & 22 & 16 & 0 & 4 \\
\hline
\end{tabular}
\caption{\label{tab:ld_data}Pacing strategies for the subset of IAAF-ratified world records for which halfway times are recorded by the IAAF.\cite{IAAF}}
\end{center}
\end{table}

\begin{figure}[h]
\begin{center}
\includegraphics{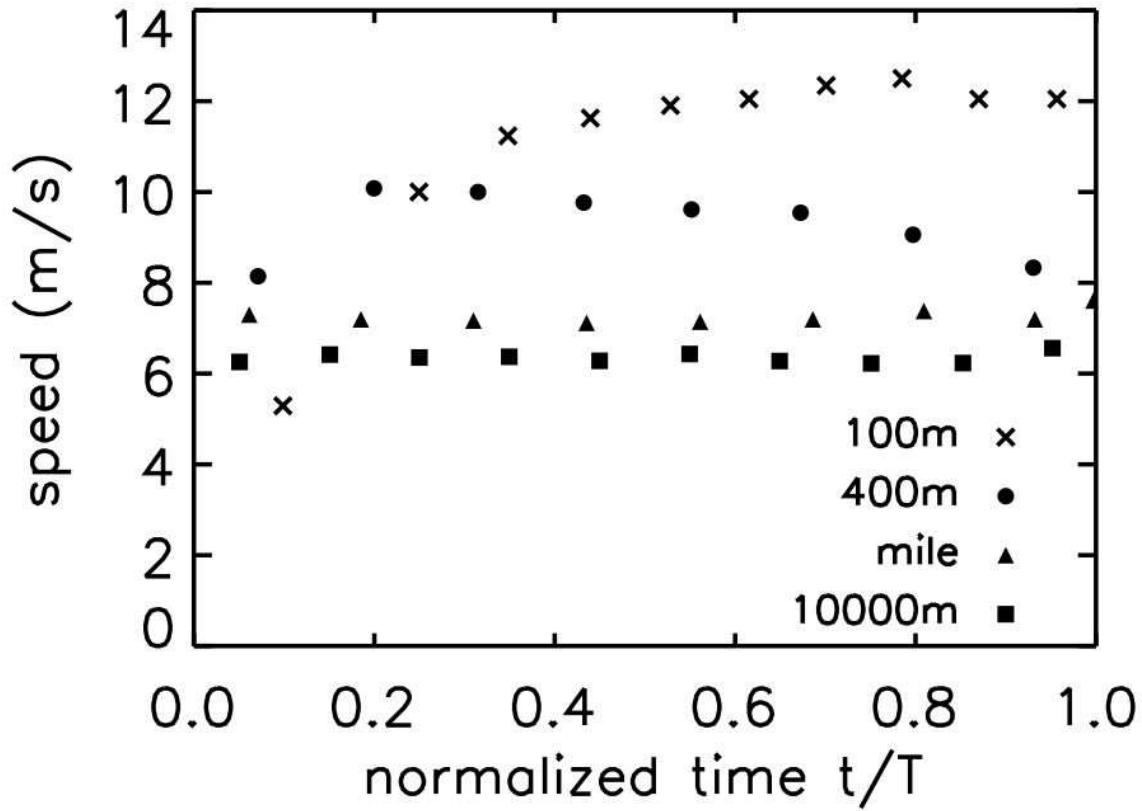}
\caption{\label{fig:pacing_strategies}Observed pacing strategies of current world record races in the men's 100 m, 400 m, mile, and 10000 m.}
\end{center}
\end{figure}

Recent improvements to videocamera technology have increased the number and accuracy of the midway split times available for many world-record performances.  Figure~\ref{fig:pacing_strategies} shows the pacing strategies for four current men's world-record performances for which multiple midway split times are available:  100 m, Usain Bolt, 9.58 s; 400 m, Michael Johnson, 43.18 s; mile, Hicham El Guerrouj, 3:43.13; 10000 m, Kenenisa Bekele, 26:17.53.  
For each of these distances the men's world record was set more recently than the women's world record, and split times (from Ref. \raisebox{-1.2ex}{\Large\cite{IAAF}\normalsize}) are more numerous, more accurate, or both.  The pacing strategy for the 100 m shows that it took the runner at least half the race to reach top speed, followed by a slight deceleration in the last 20$\%$ of the race.  The pacing strategies for the mile and 10000 m show a relatively constant pace through the whole race.  The pacing strategy for the 400 m, unlike the other races, shows that maximum speed was reached quite early in race, followed by continuous deceleration.

\section{Two Existing Mathematical Models of Optimum Race Pacing}

Keller~\cite{Keller} derived the optimal race strategy for one-lap and half-lap races, via the calculus of variations, from a model based on Newton's second law.  Keller wrote the equation of motion as

\begin{equation}
\label{eq:Keller}
\frac{dv}{dt}+\frac{v}{\tau }=f(t) ,
\end{equation}
where $v$ is the runner's speed as a function of time $t$, $\tau $ is a constant characterizing the resistance to running, assumed to be proportional to running speed, and $f(t)\leq F$ is the propulsive force per unit mass.  Empirical knowledge of human exercise physiology is expressed in the assumed relation between propulsive force and energy supply,
\begin{equation}
\label{eq:Keller2}
\frac{dE}{dt}=\sigma - fv ,
\end{equation}
where $E$ represents the runner's energy supply, which has a finite initial value $E_0$,  and is replenished at constant rate $\sigma $.  In spite of this replenishment, the energy supply reaches zero at the end of the race.  
$\tau , \sigma , E_0,$ and $F$ are found by comparing the optimal race times to the existing world records for 22 race distances ranging from 50 yd to 10000 m.  

For all race distances, this model predicts that the second half of the race will be faster than the first.  For race distances shorter than a critical distance $D_c=291$ m the optimal strategy was found to be for the runner to run at maximum propulsive force for the whole race.  The model correctly predicts that runners in 100 m and 200 m will run negative splits, and convincingly attributes this to the need to overcome inertia at the start.  For distances longer than $D_c$, the optimal strategy was to run at maximum propulsive force for one to two seconds at the start, followed by a constant speed for most of the race.  This pacing strategy is quite similar to the constant-speed pacing observed in track races of distances of a mile~\cite{Noakes_2009} and longer,~\cite{Tucker_2006} for which the initial acceleration phase lasts for a negligible fraction of the race.   

Ward-Smith~\cite{WardSmith_1985} pursued a model based on the first law of thermodynamics, and on more detailed knowledge of the bioenergetics of human metabolism.   Energy, initially in the form of chemical stored energy, is modeled as coming from two mechanisms:  an aerobic mechanism that has a maximum power output R, which can be maintained indefinitely, and an anaerobic mechanism, which decays exponentially.  The runner always uses the maximum available power.  Energy expenditures are modeled as:  energy converted to kinetic energy of the center of mass, energy needed to overcome air drag, and energy degraded to heat during the act of running.  The power balance equation is then written
\begin{equation}
\label{eq:WardSmith}
\lambda S_0 e^{-\lambda t} + R = Av + \frac{1}{2}\rho v^{3} S C_D + mv \frac{dv}{dt}\, \, ,
\end{equation}
and is numerically integrated to give $v(t)$.  Here $m$ is the runner's mass, $S$ the runner's effective surface area, and $\rho $ the ambient air density. The remaining five parameters of the model are deduced from comparing the results of the model to finishing times of runners in the Olympic Games, aided by comparison with results of various treadmill tests reported in the literature.  This model does an excellent job predicting in detail the pacing strategy used by runners in the 100 m, including the small but apparently unavoidable deceleration in the final 20 m or so of the race.  

How do these models compare to observed pacing strategies?  Midway split times~\cite{IAAF} for the current men's 400 m world record (43.18 s, Michael Johnson, 1999) are shown in the second column of Table~\ref{tab:400mcomparison}.  A velocity decrement of approximately 20\% of maximum velocity is observed during the latter three-fourths of the race, typical of race strategies observed during 400 m races by world-class, national-class, and regional-class athletes.~\cite{Hanon_2009}  The predictions of Keller's energy depletion (ED) model, and Ward-Smith's maximum available power (MAP) model, are shown in the third and fourth columns of Table~\ref{tab:400mcomparison}.  Midway split times for the current men's 800 m world record (101.01 s, David Rudisha, 2010) are shown in the second column of Table~\ref{tab:800mcomparison}.  The velocity decrement during the later stages of the race, approximately 10\%, is again typical of what is observed in high-level competition (Gajer et al.,~\cite{Gajer_2000} as shown in Fig. 1 of Hanon et al.~\cite{Hanon_2002}). 
\begin{table}[h]
\begin{center}
\begin{tabular}{|c||c||c|c|c|}
\hline
Elapsed & Actual  & Splits in & Splits in & Splits in  \\ [-1ex]
Distance & Elapsed & ED Model & MAP model & X-factor \\[-1ex]
  & Time (s) & (s) & (s) & Model (s) \\[-1ex]
   & & & & \\[-2ex]
 \hline 
 50 m & 6.14 & 6.03 & 5.52 & 6.16 \\
 \hline
 100 m & 11.10 & 11.30 & 10.09 & 11.05 \\
 \hline
 150 m & 16.10 & 16.57 & 14.88 & 16.06 \\
\hline 
 200 m & 21.22 & 21.84 & 19.94 & 21.20 \\
\hline
 250 m & 26.42 &  27.11 & 25.30 & 26.47 \\
\hline
 300 m & 31.66 & 32.38 & 30.96 & 31.88 \\
\hline
 350 m & 37.18 & 37.87 & 36.92 & 37.46 \\
\hline
 400 m & 43.18 & 43.18 & 43.18 & 43.18 \\
\hline
\end{tabular}
\caption{\label{tab:400mcomparison}Comparison of split times in the men's 400 m world record with split times predicted by various models.}
\end{center}
\end{table}
\begin{table}[h]
\begin{center}
\begin{tabular}{|c||c||c|c|c|}
\hline
Elapsed & Actual  & Splits in & Splits in & Splits in  \\ [-1ex]
Distance & Elapsed & ED Model & MAP model & X-factor \\[-1ex]
  & Time (s) & (s) & (s) & Model (s) \\[-1ex]
   & & & & \\[-2ex]
\hline
 
 200 m & 24.5 & 25.70 & 19.96 & 24.49 \\
 \hline
 400 m & 48.9 & 50.69 & 43.06 & 48.82 \\
\hline
 600 m & 74.59 & 75.68 & 70.54 & 74.29 \\
\hline
 800 m & 101.01 & 101.01 & 101.01 & 101.01 \\
\hline
\end{tabular}
\caption{\label{tab:800mcomparison}Comparison of split times in the men's 800 m world record with split times predicted by various models.}
\end{center}
\end{table}

The difference between the world-record pacing strategy and each of the two models is significant.   
Experienced professional runners must have good pace judgement:  it would not be unreasonable to expect a professional 400 m runner to be able to run the first 200 m of a 400 m race within 0.25 s of a predetermined time.~\cite{Rock_2}  The ED model predicts a second half of the race faster than the first, which has been observed in 0 of 34 one-lap, and 3 of 40 two-lap world-record races.  The MAP model dictates that a runner attempting to set a personal best in a race set personal bests at all intervening distances en route, which is never observed for trained runners in races 400 m or longer.  A new theoretical idea is needed if useful guidance is to be given to 400 m and 800 m runners regarding pacing strategy.  

\section{Model of anaerobic metabolism}

Elite athletes in 400 m and 800 m report end-of-race sensations that likely sound familiar to average athletes:  ``burning legs", in which the leg muscles are on the verge of ceasing to function.\cite{Rock,Hamilton}  It has long been known that muscle fibers that are forced to work in anaerobic conditions (without a supply of oxygen) quickly cease to function.\cite{Meyerhof,Hill}   The experience of the elite athletes suggests that race performance in the 400 m and 800 m races is limited by some change, localized to the working muscles in the legs, related to anaerobic metabolism.

\begin{figure}[h]
\begin{center}
\includegraphics{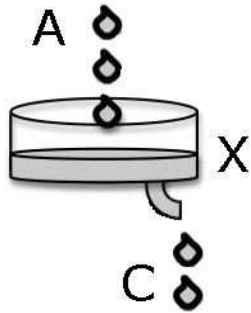}
\caption{\label{fig:phys_mod}Physical model representing anaerobic metabolism.}
\end{center}
\end{figure}

A toy model representing anaerobic metabolism in working muscle is shown in Fig.~\ref{fig:phys_mod}.  
Process $A$ represents the production of X-factor, an as-yet-unspecified metabolite which is associated with the cessation of function of muscles working in anaerobic conditions.  $A(v)$, the rate of production of X-factor, is a function of running speed $v$ that satisfies $d^{2}A/dv^{2}>0$.  $X$ represents the amount of X-factor accumulated in working muscle.  There is a maximum value $X_{F}$, such that when $X$ reaches this value in a working muscle, that muscle ceases to function.  Process $C$ represents the removal or disappearance of X-factor from working muscle.

The key assumption of the model (see Sec.~\ref{sec:Discussion}) is that the rate of removal or disappearance of X-factor is proportional to the amount of X-factor accumulated.  With this assumption, the amount $X(t)$ of X-factor in working muscle is described by the differential equation
\begin{equation}
\label{eq:x_diff_eq}
\frac{dX(t)}{dt}=A(v)-CX(t) ,
\end{equation}
where the running speed $v$ is itself a function of time, and $C$ is a constant.  The runner starts the race at $x=0, t=0$ with $X=X_{0}$, the amount of X-factor in resting muscle.  If the race is to be run optimally, the runner finishes the race with $X=X_{F}$.

With this model, determination of optimal race strategy is perfectly suited to treatment by the calculus of variations, leading to the Euler-Lagrange equation.  Rather than consider the problem that presents itself to the runner--how to minimize the time needed to cover a given distance d--it is analytically more convenient to consider the related problem of how to maximize the distance covered in a given time $T$.  This leads to the equation
\begin{equation}
\label{eq:EL2}
x=\! \int_{start}^{finish} dx = \! \int_{0}^{T} v\, dt  ,
\end{equation}
where $v$ is now considered a function of $X$ and $\dot{X}=\partial X/\partial t$.  Using Eq.~(\ref{eq:EL2}), the Euler-Lagrange equation is $\partial v/\partial X = d/dt\, [\partial v/\partial \dot{X}]$.  Since $X$ and $\dot{X}$ enter only in the combination $\dot{X}+C{X}\equiv y$, the Euler-Lagrange equation can always, regardless of the functional form of $A(v)$, be written
\begin{equation}
C\frac{dv}{dy}=\frac{d}{dt} \frac{dv}{dy},
\end{equation}
and then integrated once to give
\begin{equation}
\label{eq:y_diff_eq}
\frac{dv}{dy}=ke^{Ct},
\end{equation}
where $k$ is a constant of integration.    Rewriting Eq.~(\ref{eq:x_diff_eq}) as $y=A(v)$ and inverting gives $v=A^{-1}(y)$, from which the left-hand-side of Eq.~(\ref{eq:y_diff_eq})  follows, and is rewritten in terms of $v$.  This gives the optimal $v(t)$ in terms of the unknown constant $k$, which is then be found by rearranging Eq.~\ref{eq:x_diff_eq}, multiplying both sides by the integrating factor $e^{Ct}$, and integrating from $0$ to $T$, to give
\begin{equation}
\label{eq:x_from_a}
X_{F}e^{cT}-X_{0}= \! \int_{0}^{T} A\{v(t;k)\} e^{Ct} dt\, .
\end{equation}

\section{Example}

Consider $A(v)=A_{0}(v/v_{0})^{n}+CX_{0}$, where $A_{0}$, $v_{0}$ and $n$ are free parameters.  Equation~(\ref{eq:x_diff_eq}) may be rearranged to give $v$ in terms of $y$,
\begin{equation}
\label{eq:v_from_y}
v=v_{0}\left(\frac{\dot{X}+CX-CX_{0}}{A_{0}}\right)^{1/n}=v_{0}\left(\frac{y-CX_{0}}{A_{0}}\right)^{1/n},
\end{equation}
from which the left side of Eq.~(\ref{eq:y_diff_eq}) is 
\begin{equation}
\frac{dv}{dy}=\frac{v_{0}}{nA_{0}}\left(\frac{y-CX_{0}}{A_{0}}\right)^{\frac{1-n}{n}}=\frac{v^{n}_{0}}{nA_{0}}\,\,  v^{1-n},
\end{equation}
and, rearranging Eq.~(\ref{eq:y_diff_eq}), $v$ is found in terms of the unknown constant $k$:
\begin{equation}
v=\left(\frac{nA_{0}k}{v_{0}^n}\right)^{\frac{1}{1-n}}e^{-Ct/(n-1)}.
\end{equation}
$k$ is found by inserting this into Eq.~(\ref{eq:x_from_a}), and integrating once.  The optimal time histories for $v$ and $X$ are
\begin{equation}
\label{eq:v_power_law}
v=v_{0}\left(\frac{C}{(n-1)A_{0}}\right)^{1/n}\left(\frac{(X_{F}-X_{0})e^{CT}}{1-e^{\frac{-CT}{n-1}}}\right)^{1/n}e^{-Ct/(n-1)},
\end{equation}

\begin{equation}
X(t)=X_{0}+\frac{(X_{F}-X_{0})e^{CT}}{1-e^{\frac{-CT}{n-1}}}\left(e^{-Ct} - e^{-nCt/(n-1)}\right) \, \, .
\end{equation}

For $n>1$, the optimal speed as a function of time is monotonically decreasing--``positive splits".  Only for $n<1$ is the optimal velocity monotonically increasing with time--``negative splits".  This implies that for any $A$ for which $d^{2}A/dv^{2} > 0$ for all $v$, if an optimal pacing strategy exists, that optimal pacing strategy requires positive splits.  The race may be imagined as divided into segments, during each of which the speed is restricted to sufficiently small range that $A(v)$ over that range may be approximated as a power law, with appropriate $n>1$.  Then the speed during each segment, and therefore the speed during the whole race, will be a monotonically decreasing function of time.   

\section{Accounting for the Start}

One obvious drawback of the optimal $v(t)$ found above is that the optimal $v(t=0) \neq 0$, which contradicts the fact that runners start from rest.  It might be thought that one way around this would be to suppose that X-factor production also depends on acceleration, and write $A=A(v,\dot{v})$.  One might suppose that in an increment of time $\Delta t$, an increment of chemical stored energy $\Delta E$ is converted into an increment of work $\Delta W$, and an increment of heat $\Delta Q$; and that this is accompanied by the generation of an increment of X-factor $\Delta X$.  Since $\Delta E=\Delta W + \Delta Q$, if one further supposes that $\Delta Q, \Delta X \propto \Delta E$, and that $\Delta W$ is to be identified with the change in mechanical energy, then for races on level ground one is led to the ansatz
\begin{equation}
\Delta X \propto \Delta W = \frac{d}{dt}\, \left[ \frac{1}{2}\, mv^{2}\right] \Delta t = mv\, \dot{v} \, \Delta t, \\
\implies A=A_{0}v\, \dot{v}\ .
\end{equation}

This ansatz is a dead end.  Mathematically, one can see this by treating the equation $\dot{X}+CX-A(v,\dot{v})=0$ as a constraint, and using the method of Lagrange multipliers, 
\begin{equation}
x=\! \int_{start}^{finish} dx = \! \int_{0}^{T}\left[ v + \lambda (t) \left( \dot{X}+CX-A(v,\dot{v}) \right)\right] \, dt \, ,
\end{equation}
which leads to two Euler-Lagrange equations.  The equation involving derivatives with respect to $X$ and $\dot{X}$, $\partial L/\partial X- d/dt\, [\partial L/\partial \dot{X}]=0$, leads to $\lambda (t) = k e^{Ct}$.  The terms involving $X$ and $\dot{X}$ then form a total time derivative, and therefore do not contribute to the variation of the integrand.  However, any term of the form $A=A_{1}(v)\dot{v}$ may be written 
\begin{equation}
A= A_{1}(v)\dot{v}+CA_{2}(v)-CA_{2}(v), 
\end{equation}
where $A_{2}(v)$ is chosen so that $d/dv [A_{2}(v)]=A_{1}(v)$.  Then the term involving $\dot{v}$ will also be part of a total time derivative, and will also not contribute the variation.  Thus the model $\dot{X}+CX=A$ possesses a gauge freedom, so that replacing $A_{0}(v)\rightarrow A_{0}(v)+CA_{2}(v)+\dot{v}\, d/dv[A_{2}(v)]$ has no effect on the ``equation of motion", the optimal $v(t)$.  

This mathematical result is a reflection of the physical fact that the acceleration phase at the start of the race is not limited by any type of burning sensation, localized or otherwise, but instead by the runner's inertia, as implied by both the ED and MAP models.  Treating this finite limit to acceleration as a constraint, and retaining $A=A(v)$, it may be proven that the best possible pacing strategy has two phases:  an acceleration phase, limited by the runner's inertia, with monotonically increasing velocity, followed by a deceleration phase, limited by X-factor accumulation, with monotonically decreasing velocity.

\begin{figure}[h]
\begin{center}
\includegraphics{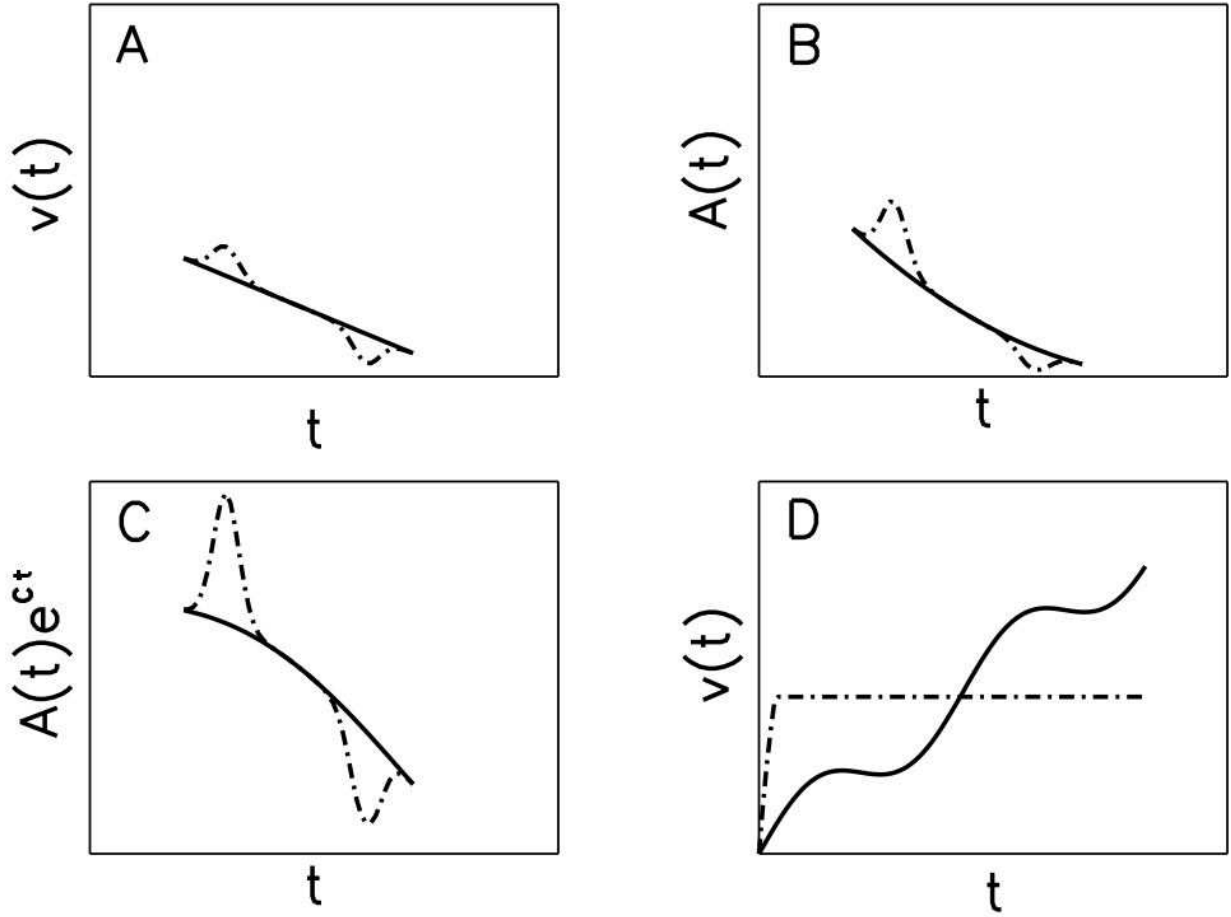}
\caption{\label{fig:optimum_proof}A. Two pacing strategies that cover the same distance (areas under the two curves are the same):  an ``original" strategy (solid line) and a ``perturbed" strategy (dot-dashed line).  B.  X-factor generated during the course of the race by the perturbed pacing strategy (area under dashed curve) is more than the X-factor generated by the original pacing strategy (area under solid curve).  C.  X-factor accumulated in the working muscle by the end of the race may or may not be larger for the perturbed than the original strategy, depending on the details of the perturbation (area under dashed curve may or may not be larger than area under solid curve).  D.  A case in which perturbed strategy is superior to original strategy.}
\end{center}
\end{figure}

A graphical outline of the proof is as follows.  Consider two trial pacing strategies as shown in Fig.~\ref{fig:optimum_proof}A:  ``original" pacing (solid line) and ``perturbed" pacing (dashed line).  
The areas under the two curves are the same (the same distance is covered).  The function $A(t)$ governs how much X-factor is generated (as opposed to accumulated in the working muscle).  More X-factor is generated by the perturbed strategy than by the original strategy (Fig.~\ref{fig:optimum_proof}B), because the positive perturbation to $v$ happens at a higher $v$ than the negative perturbation to $v$, and $d^{2}A/dv^{2} > 0$ for all $v$.  However, the more X-factor is present, the more can be removed.  As may be seen from Eq.~\ref{eq:x_from_a}, $X_F$ increases as $ \! \int_{0}^{T} A(t)e^{Ct} dt$ increases.  The relative $X_F$ produced by the two pacing strategies may be compared by comparing the areas under the two curves in Fig.~\ref{fig:optimum_proof}C.  The perturbed strategy is advantaged because the positive perturbation to $v$ happens early.  The superior strategy is the one that results in the lowest $X_F$.  Which strategy is superior depends on the details of the perturbation.  
 
Taking the perturbations as of an infinitesimal nature, occurring at times $t_1$ and $t_1+\delta $, where $\delta $ is finite, the perturbed strategy is better than the original strategy if
\begin{equation}
\label{eq:optimum_proof}
\frac{\left. {\frac{dA(t)}{dt}}\right| _{t_{1}}}{\left. {\frac{dA(t)}{dt}}\right| _{t_{1}+\delta }} < \frac{e^{C(t_{1} +\delta )}}{e^{Ct_{1}}} = e^{C\delta }.
\end{equation}
Using  $A(v)=A_{0}(v/v_{0})^{n}$ as an example, one finds that the perturbed and original strategies are equally good if $v(t_{1}+\delta ) = v(t_{1}) e^{-C\delta /(n-1)}$, which is the same time dependence found from the Euler-Lagrange equation (Eq.~\ref{eq:v_power_law}).  

Similar reasoning shows that a pacing strategy that does not start out at maximum possible acceleration is inferior to one that does.  Figure~\ref{fig:optimum_proof}D shows an arbitrary pacing strategy (solid line) which is inferior to a pacing strategy that starts at maximum acceleration, then continues at a constant speed (dashed line).  

\begin{figure}[h]
\begin{center}
\includegraphics{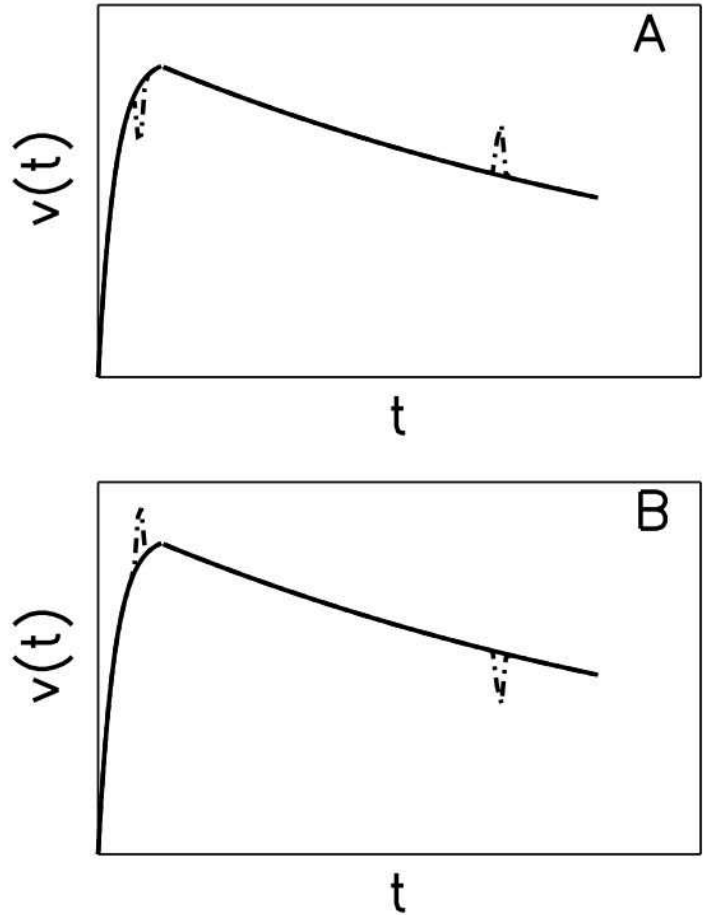}
\caption{\label{fig:optimum_proof2}Two possible perturbations to the optimal pacing strategy of maximum acceleration followed by optimal deceleration.  A.  Perturbations of this type are inferior to the optimal pacing strategy.  B.  Perturbations of this type are not possible.}
\end{center}
\end{figure}

By applying a sequence of infinitesimal perturbations of the type described above, it may be shown that no possible pacing strategy is superior to the strategy of accelerating with maximum possible acceleration until such time as the Euler-Lagrange optimum pacing connecting one's current $(t,X)$ and the finish $(T,X_{F})$ does not require an instantaneous jump in speed (assumed to be forbidden), and continuing at the Euler-Lagrange optimum pacing until the finish.  This is the best possible pacing strategy.  It is shown in Fig.~\ref{fig:optimum_proof2}, along with two possible perturbations.  The perturbation shown by the dashed curve in Fig.~\ref{fig:optimum_proof2}A fails to satisfy Eq.~\ref{eq:optimum_proof}, and results in a higher $X_{F}$ at the finish than the best possible pacing.  The perturbation shown in Fig.~\ref{fig:optimum_proof2}B is not possible because it exceeds the maximum possible acceleration.   

\section{Comparison of Models}
Attempts were made to reproduce the split times recorded during the men's 400 m and 800 m world record races.  During the men's 400 m world record race (duration 43.18 s), the athletes started from starting blocks equipped with electronic force sensors.  The time elapsed between the start of the race and significant force generation by Michael Johnson was recorded to be 0.150 s.\cite{IAAF_web}  The ED model, MAP model, and X-factor model were therefore made to represent an athlete who remained motionless at the start for 0.150 s, and crossed the finish line at 43.18 s.  

Split times for the ED and MAP models were computed by using the published model parameters to generate pacing strategies.  This resulted in races finished in 43.27 s for the ED model, and 44.31 s for the MAP model.  Split times for the ED model were then multiplied by 43.03/43.27 and split times for the MAP model by 43.03/44.31; finally 0.15 s of motionlessness was added at the start. 

For the X-factor accumulation model, X-factor was assumed to accumulate as $A(v)=A_{0}(v/v_{0})^{3}+CX_{0}$, from a value at rest of $X_{0}$ to a value at the finish of $X_{F}$; the acceleration phase was described as $v(t)=v_{max}(1-e^{-c_{2}t})$.  Writing the units of X-factor as [X], parameter values were $A_0=0.05$ [X] /s, $v_{0}=6.5$ m/s, $C=0.01$ s$^{-1}$, $X_{0}= 1.0$ [X], $X_f = 6.27$ [X], $T=43.03$ s, $v_{max}=11.7$ m/s, $c_2=0.60$ s$^{-1}$.  

A comparison of the split times predicted by the three models with the actual split times in the 400 m world record race is shown in Table~\ref{tab:400mcomparison}.  For the $350$ m segment time, the MAP model is closer to the actual time by 0.02 s than the X-factor model is, but for all the other segment times the X-factor model is the closest of the three models.  

No blocks were used during the men's 800 m world record race.  The motionless period at the start was estimated to be 0.3 s, and 200 m split time was determined to be 24.5 s, from video analysis by the author; 400 m split time was given by Track and Field News as 48.9 s;~\cite{TFN_Rudisha} 600 m split time and finish time are from Ref. \raisebox{-1.2ex}{\Large\cite{IAAF}\normalsize}.  

Split times for the ED and MAP models were again computed from published model parameters, resulting in races finished in 105.95 s (ED model) and  104.38 s (MAP model), and then scaled to the world record finishing time as before.  The X-factor model parameters were $A(v)=A_{0}(v/v_{0})^{6}+CX_{0}$ with the acceleration phase $v(t)$ as for the 400 m case, $A_0=0.05$ [X]/s, $v_{0}=7.337$ m/s, $C=0.0092$ s$^{-1}$, $X_{0}= 1.0$ [X], $X_f = 6.27$ [X], $T=100.71$ s, $v_{max}=11.3$ m/s, $c_2=0.63$ s$^{-1}$.  Results are shown in Table~\ref{tab:800mcomparison}.  The X-factor accumulation model predicts each of the split times much better than either of the other two models.  

It must be stressed that the mathematical functions used here for X-factor accumulation rate $A(v)$, while plausible, were chosen for analytic convenience and are not the result of a theory of X-factor accumulation.  However, the result that the optimal pacing strategy for 400 m and 800 m races is to start at a pace faster than can be sustained until the end of the race is robust, and not dependent on the particular function used for X-factor accumulation rate.  

\section{Discussion}
\label{sec:Discussion}

The traditional view is that muscles cease to function during high-intensity, anaerobic exercise due to a buildup of $H^{+}$ ions, which presumably leads to deactivation of enzymes needed for energy production.  The strongest support for this comes from the many studies of athletes performing various high-intensity exercises to voluntary exhaustion.  For example, Sahlin et al.~\cite{Sahlin_1978a} found via biopsy that the intracellular pH in the working muscle (quadriceps femoris) in exhausted cyclists had dropped from a baseline of 7.0 to 6.4, while Costill et al.~\cite{Costill_1983} found via biopsy that the pH in the gastrocnemius, after an all-out 400 m sprint, had dropped from a baseline of 7.03 to 6.63, and theorized that runners, needing to maintain upright posture, could not tolerate so low a pH as cyclists.  Furthermore, ingestion of bicarbonate (``soda loading") improves an athlete's capacity for intense anaerobic exercise, perhaps due to increased extracellular buffering of $H^+$.~\cite{Wilkes_1983}

This traditional view is challenged by a variety of recent research results.  For example, the pH of working muscles does not monotonically decrease during anaerobic exercise:  upon initiation of exercise, there is a transient increase.~\cite{Kemp_2001,Walter_1999}  This transient alkalosis is presumed to result from production of ATP by splitting of phosphocreatine, a process which consumes $H^{+}$ ions.~\cite{Hultman_1980}  An alternative explanation that has been advanced for the cessation of function of the working muscles in 400 m runners is the depletion of phosphocreatine in the working muscles.~\cite{Hirvonen_1992}  For a review of literature challenging the traditional view, see Allen et al.~\cite{Allen_2008}

Since no single biochemical species may be unequivocally identified as the X-factor, it is suggested that the X-factor is best thought of as entropy.  In making this suggestion, no attempt is made to favor one or another of the competing biochemical hypotheses concerning the cause of cessation of muscle function in high-intensity exercise, or to suggest that inadequate heat dissipation is the cause.  Rather, the suggestion is made that a muscle at rest may be considered from the point of view of irreversible thermodynamics as a non-equilibrium stationary state, in which entropy-producing processes are sustained at constant levels by continual fluxes of energy and matter, and efflux of entropy.  See Chapter 1 of  Ref. \raisebox{-1.2ex}{\Large\cite{Yourgrau}\normalsize} for background to this perspective.  

If the X-factor represents entropy, then $X$ represents entropy density, and Eq.~(\ref{eq:x_diff_eq}) is to be interpreted as the continuity equation for entropy regarding an entropy-rich region (a working muscle), of entropy density $X$, enclosed by a boundary that is resistant to the efflux of entropy, and surrounded by an entropy-poor region.  Equation~(\ref{eq:x_diff_eq}) results by assuming the amount of entropy that crosses the barrier is insufficient to significantly raise the entropy density in the external region.  The quantity $C$ describes the rate of diffusive efflux of entropy from the entropy-rich region.  The function $A(v)$, describing the rate of generation of entropy as a function of running speed, is expected to grow at least as fast with $v$ as the rate of consumption of energy.  Therefore it should satisfy $d^2A/dv^2 > 0$, since the runner encounters velocity-dependent frictional forces, such as air resistance.

With this interpretation, the difference in race tactics employed in 400 m and 800 m races from the pacing strategies employed in shorter and longer races is understood as follows.  Shorter races do not lead to high enough entropy density to cause loss of function in muscles.  For longer races, the advantages of starting fast are blunted because the entropy density outside the working muscle (for example, in the blood) can no longer be considered small compared to the entropy density inside the working muscle.  Only in the 400 m and 800 m races is the entropy efflux term important enough to recommend positive-split pacing.

The X-factor accumulation model suggests that runners training for 400 m and 800 m races should perform exercises that maximize the entropy density gradient across the boundaries of the working muscles in the legs.  This might be expected to increase the abilities of the working muscles to rid themselves of entropy by diffusion.  It seems apparent that the best way to maximize the entropy density gradient is to ask muscles to perform work in the absence of blood flow.  It has long been known that sustained muscular contraction reduces blood flow through the contracting muscle,~\cite{Barcroft_1939} and there seems agreement that cessation of blood flow can be accomplished by sustained contraction at greater than 60-70\% of maximum voluntary force.~\cite{Gaffney_1990}  This is within the range of sustained muscle force required during certain weightlifting exercises, such as squats, favored by sprinters.  It may be speculated that, in addition to building muscular strength, such weightlifting exercises aid performance in the 400 m and 800 m races by increasing the abilities of the working muscles in the legs to rid themselves of entropy by diffusion.  

\section{Conclusions}

Runners in 400 m and 800 m races are often urged to attempt to run the second half of the race at the same pace as the first (``even splits"), on the grounds that this is the most economical distribution of energy, as no doubt it is.   However, almost no successful runners in 400 m and 800 m races actually do this, and the burden of this paper has been to try to understand why.  A simple model of entropy accumulation suggests that for any reasonable relation between entropy generation and speed, the optimal race strategy is to run the first half of the race faster than the second.  Such a strategy causes the entropy density gradient across the boundary of the working muscle to take on a large value early in the race, which increases the entropy efflux.  That this is in fact the optimal race strategy is proven by the overwhelming preponderance of positive-split races in the set of world records for one-lap and two-lap races.  Since there are many other factors that athletes encounter in races, such as the race tactics of other athletes, as well as psychological variables, this analysis may not justify advising an athlete who prefers to run even-split, or even negative-split races, to adopt a different pacing strategy.  However, one should not, on the basis of energy arguments, discourage an athlete who prefers to run positive-split 400 m or 800 m races from doing so, because it may be that the athlete knows best.  

\begin{acknowledgments}
Thanks to Susan Coppersmith, Dushko Kuzmanovski, and Andrew Rock for encouragement and comment.  
\end{acknowledgments}

\end{document}